%% file: main.tex
\documentclass[lettersize,journal,compsoc]{IEEEtran}
\usepackage{amsmath,amsfonts}
\usepackage{algorithmic}
\usepackage[ruled,vlined]{algorithm2e}
\usepackage{array}
\usepackage[caption=false,font=normalsize,labelfont=sf,textfont=sf]{subfig}
\usepackage{textcomp}
\usepackage{stfloats}
\usepackage{url}
\usepackage{verbatim}
\usepackage{graphicx}
\usepackage{cite}
\usepackage{tikz}
\usepackage{jabbrv}
\usepackage{amssymb}   
\usepackage{pifont}    
\usepackage{graphicx}
\usepackage{booktabs}
\usepackage{multirow}
\usepackage{xcolor}
\usepackage{tabularx}
\usepackage{makecell}
\usepackage{float}
\usepackage{placeins}
\usepackage{ragged2e}
\usepackage{subcaption}

\newcommand{\circnum}[1]{%
  \tikz[baseline=(char.base)]{
    \node[
      shape=circle,
      draw=black,
      fill=black,
      inner sep=0pt,
      minimum size=10pt
    ] (char) {\textcolor{white}{\scriptsize #1}};
  }%
}

\newcommand*\circled[1]{%
  \tikz[baseline=(char.base)]{
    \node[shape=circle,draw,inner sep=1.5pt] (char) {\scriptsize #1};
  }
}

\usepackage{soul,color,xcolor}      
\definecolor{myColor}{rgb}{0,0,0}        
\makeatletter
\newcommand*{\revise}{\@ifnextchar\bgroup{\revise@}{\color{myColor}}}
\newcommand*{\revise@}[1]{{\textcolor{myColor}{#1}}}
\makeatother

\newcolumntype{?}{!{\vrule width 1pt}}

\usepackage{etoolbox}

\makeatletter
\pretocmd{\@bibitem}{%
  \ifstrequal{#1}{abera2016c}{\color{myColor}}{
  \ifstrequal{#1}{zhou2020silhouette}{\color{myColor}}{
  \ifstrequal{#1}{rattanavipanon2025slapp}{\color{myColor}}{
  \color{black}}}}
}{}{}

\pretocmd{\@lbibitem}{%
  \ifstrequal{#2}{abera2016c}{\color{myColor}}{%
  \ifstrequal{#2}{zhou2020silhouette}{\color{myColor}}{%
  \ifstrequal{#2}{rattanavipanon2025slapp}{\color{myColor}}{%
  \color{black}}}}%
}{}{}
\makeatother

\begin{document}

\title{PS-UIE: Privilege-Separated Integrity Enforcement for User-Space Executable Objects in Confidential VMs}

\author{Jingkai Mao, Xiaolin Chang 
\IEEEcompsocitemizethanks{
\IEEEcompsocthanksitem Jingkai Mao and Xiaolin Chang are with the Beijing Key Laboratory of Security and Privacy in Intelligent Transportation, Beijing Jiaotong University, P.R.China. (e-mail: \{23111143, xlchang\}@bjtu.edu.cn)
}
}

\markboth{Journal of \LaTeX\ Class Files,~Vol.~14, No.~8, August~2021}%
{Shell \MakeLowercase{\textit{et al.}}: A Sample Article Using IEEEtran.cls for IEEE Journals}

\IEEEpubid{}

\input{tex/0-abs}
\input{tex/1-intro}
\input{tex/2-background_related}

\input{tex/3-system}
\input{tex/4-design}

\input{tex/5-implementation}
\input{tex/6-evaluation}
\input{tex/8-conclusion}
\input{tex/9-references}
\input{tex/10-biography}

\end{document}

%% file: tex/0-abs.tex
\IEEEtitleabstractindextext{
\begin{abstract}
\justifying
Confidential Virtual Machines (CVMs), such as AMD SEV-SNP, enable cloud tenants to run security-sensitive workloads, but tenants can rely on the execution of these workloads only when they can trust the CVM. This trust requires continuous integrity assurance from CVM launch to the current runtime state, including initial trust establishment at launch and subsequent runtime integrity assurance. Existing works help establish launch-time trust and protect parts of runtime integrity, but they do not fully address the integrity of file-backed user-space executable objects, such as main executables, program interpreters, and dynamically loaded shared objects, that may be loaded or mapped dynamically during execution inside CVMs.

In this paper, we propose Privilege-Separated User-space Integrity Enforcement (PS-UIE), an approach for enforcing the integrity of user-space executable objects inside AMD SEV-SNP-based CVMs. PS-UIE consists of a privilege-separated architecture and three mechanisms. The architecture separates the authority for integrity measurement and enforcement from the measured targets by placing it in a higher-privileged protected domain. Built on this architecture, PS-UIE provides \textit{policy lifecycle management}, \textit{execution-time integrity enforcement}, and \textit{evidence export and verification} mechanisms. It enables policy-controlled integrity measurement and enforcement for user-space executable objects and generates verifiable runtime evidence. We implement PS-UIE on an AMD SEV-SNP platform. The security analysis and performance evaluation show that PS-UIE enforces the integrity of user-space executable objects on the covered execute-permission grant paths and provides verifiable runtime evidence while incurring acceptable overhead.
\end{abstract}

}

\maketitle

%% file: tex/1-intro.tex
\section{Introduction}

\IEEEPARstart{C}{onfidential} Virtual Machines (CVMs), such as AMD SEV-SNP, Intel TDX, and ARM CCA, enable cloud tenants to run security-sensitive workloads inside a protected guest VM, referred to as the guest, with hardware isolation and reduced trust in cloud providers~\cite{feng2024survey}. Because such workloads often handle sensitive code or data in confidential services, data processing, and AI workloads~\cite{mo2024machine}, tenants can deploy them with confidence only when they can trust the guest. Trust in the guest requires its integrity state to remain consistent with the expected state throughout its life-cycle~\cite{ammar2025sok}.

Maintaining such life-cycle trust requires continuous integrity assurance from launch to the current runtime state, which relies on launch-time trust establishment and \textbf{runtime integrity} assurance~\cite{ammar2025sok}. At launch time, remote attestation provides the initial integrity evidence, allowing the tenant to obtain a trusted report of the guest launch state and decide whether to trust the guest before provisioning secrets~\cite{wilke2024snpguard}. However, this launch-time evidence does not guarantee runtime integrity, because software inside the guest may later load kernel modules, start user-space services and applications, and map new executable code. These runtime changes can make the current integrity state diverge from the state validated at launch, especially in a complex guest operating system (OS) where code may be dynamically built, loaded, modified, and executed~\cite{wang2025road}. 

Existing work on runtime integrity assurance inside CVMs can be broadly grouped into two lines: (i) \textit{service-level protection} and (ii) \textit{system-level integrity measurement and enforcement}. For \textit{service-level protection}, recent works use privileged in-CVM isolated service layers to protect specific security functions, services, or execution domains~\cite{narayanan2023remote,mei2024svsm,zhou2024verismo,wang2025road,ahmad2023veil,mei2024cabin,schwarz202400seven,sabanic2025confidential,zhang2025complementing,wang2025tetd}. However, these works focus on specific services or protected domains, while runtime changes in other components inside the guest may still affect the guest integrity state. For \textit{system-level integrity measurement and enforcement}, the Linux Integrity Measurement Architecture (IMA) and its extensions~\cite{sailer2004design,ima_appraisal2012,luo2019container,eckel2021userspace}, together with Trusted Execution Environment (TEE)-assisted systems~\cite{kanonov2016secure,dong2020kims,song2022tz,song2024dimac,zhao2024trusted,liu2023tzeamm,mao2025tpm2,wang2024towards,zhou2022smile,ozga2021triglav}, measure or enforce targets inside the measured system. However, these mechanisms either rely on the guest OS for integrity measurement and enforcement, which can make their results untrusted when the guest OS is compromised, or target non-CVM settings and do not directly support runtime integrity assurance inside CVMs. These limitations motivate the need for a system-level integrity authority that \textbf{performs measurement and enforcement in a more privileged protected domain, separated from the guest OS.}

\begin{figure}[!t]
\includegraphics[width=3.5in,trim=40 32 35 21,clip]{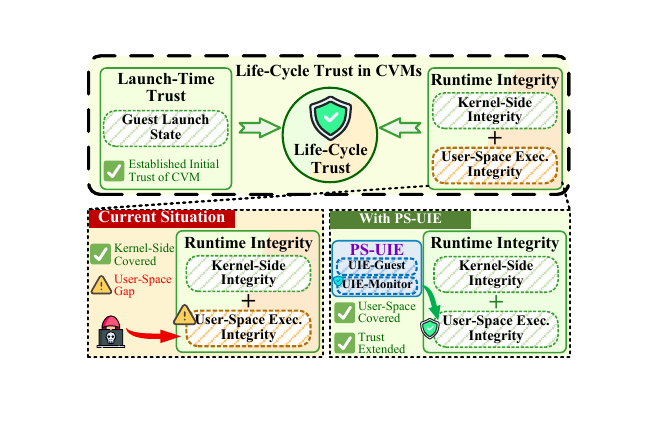} 
\caption{Trust coverage of a CVM without (bottom-left) and with PS-UIE (bottom-right). Launch-time attestation establishes initial trust, while existing runtime mechanisms cover kernel-side integrity but leave user-space executable integrity exposed. PS-UIE addresses this gap.}
\label{fig:intro}
\vspace{-10pt}
\end{figure}

Research on \textit{system-level integrity measurement and enforcement inside CVMs} remains limited. While VMPL-KMI~\cite{mei2025vmpl} implements a higher-privileged in-CVM component to protect the integrity of guest kernel modules, it does not provide integrity measurement and enforcement for file-backed user-space executable objects, including \textit{main executables}, \textit{program interpreters}, and \textit{dynamically loaded shared objects}. In this paper, we refer to these objects as \textbf{user-space executable objects} and to their integrity as \textbf{user-space executable integrity}. This integrity problem remains unresolved for establishing trust in CVMs, as shown in Fig.~\ref{fig:intro}. Addressing this gap requires measuring and enforcing user-space executable integrity inside CVMs, which raises at least \textbf{three critical challenges (C1--C3)}:

\textbf{C1: Covering Execute-Permission Paths with Privilege Separation}. User-space executable objects may be granted execute permission through multiple paths, so effective integrity enforcement must cover these paths before execute permission is granted. Given the need for privilege separation, a higher-privileged integrity authority should perform integrity measurement and enforcement~\cite{mei2025vmpl}. However, execution events originate inside the guest OS and must be captured without breaking normal execution semantics. The challenge is to cover execute-permission grant paths under privilege separation to enforce user-space executable integrity inside CVMs.
    
\textbf{C2: Enforcing User-space Executable Integrity with Acceptable Overhead}. Integrity enforcement requires measuring executable content and checking policy before execution is allowed. In a privilege-separated architecture, this process may require frequent transitions between the guest OS and a higher-privileged integrity authority, introducing visible overhead~\cite{wang2025road}. This cost is especially important for user-space executable integrity because program execution and shared-object loading occur more frequently than kernel module loading. The challenge is to keep overhead acceptable while leaving final decisions to the higher-privileged authority.

\textbf{C3: Generating Trustworthy and Verifiable Runtime Evidence}. Integrity enforcement should generate verifiable runtime evidence so that tenants can extend trust from launch time to the current runtime state. However, such evidence may be tampered with during generation, storage, or transmission. The challenge is to protect this evidence from corruption and make it verifiable by tenants without trusting the guest OS as the evidence authority.

To address these challenges, we propose a \textit{P}rivilege-\textit{S}eparated \textit{U}ser-space \textit{I}ntegrity \textit{E}nforcement (\textbf{PS-UIE}) approach that targets the user-space executable integrity gap and helps support life-cycle trust in CVMs, as shown in Fig.~\ref{fig:intro}. PS-UIE consists of a privilege-separated architecture and three mechanisms. Together, they enable integrity measurement and enforcement of user-space executable objects and generate trustworthy and verifiable runtime evidence. To the best of our knowledge, PS-UIE is the \textbf{first work} to provide integrity measurement and enforcement for user-space executable objects through a higher-privileged integrity authority in AMD SEV-SNP-based CVMs.

We implement a PS-UIE prototype on an AMD SEV-SNP platform by modifying the Linux kernel in the guest OS and the Secure VM Service Module (SVSM) in the higher-privileged service layer. We evaluate the prototype through security analysis and performance experiments using microbenchmarks and representative workloads. The results show that PS-UIE enforces user-space executable integrity on the covered execute-permission grant paths inside AMD SEV-SNP CVMs with acceptable overhead, demonstrating the feasibility of the design.

The contributions are summarized as follows:

1) \textbf{Proposing a Privilege-Separated Architecture}. We propose a privilege-separated architecture that separates execute-permission grant event collection from the integrity authority while covering execute-permission grant paths, addressing \textbf{C1}.

2) \textbf{Proposing Integrity Enforcement Mechanisms}. We propose three mechanisms to support integrity measurement and enforcement with acceptable overhead and trustworthy runtime evidence generation, addressing \textbf{C2} and \textbf{C3}.

3) \textbf{Implementing and Evaluating a Prototype}. We implement and evaluate a PS-UIE prototype to demonstrate its security, performance, and feasibility.

The rest of this paper is organized as follows. Section~\ref{sec:background_related} introduces the background and reviews related work. Section~\ref{sec:system} presents the threat model, security requirements, and system overview of PS-UIE. Sections~\ref{sec:design} and~\ref{sec:implementation} describe the design and implementation of PS-UIE, respectively. Section~\ref{sec:evaluation} reports the security analysis and performance evaluation. Finally, Section~\ref{sec:conclusion} concludes the paper.

%% file: tex/2-background_related.tex
\section{Background and Related Work}
\label{sec:background_related}

\subsection{Background}
\label{sec:background}

\subsubsection{AMD SEV-SNP}

\textbf{AMD SEV-SNP}. AMD SEV-SNP is a CVM architecture that protects a guest from an untrusted hypervisor through memory encryption and integrity protection~\cite{amdsev_snp}. It uses the Guest-Hypervisor Communication Block (GHCB) to exchange request and response state with the hypervisor through \texttt{VMGEXIT}-based communication. SEV-SNP introduces the Reverse Map Table (RMP), which is managed by the AMD Platform Security Processor (PSP) and records physical-page access permissions. During address translation, the processor checks the RMP to verify page ownership and read, write, and execute permissions. The processor raises a nested page fault if an access violates the RMP state.

\textbf{VMPL}. SEV-SNP introduces VMPLs as an in-CVM privilege separation mechanism~\cite{amdsev_snp,svsm_spec}. A CVM can use four VMPLs, and VMPL0 has the highest privilege. Each virtual CPU context is associated with a VM Save Area (VMSA) at a specific VMPL. All VMPLs share the same guest physical address space, but the RMP can assign per-VMPL read, write, and execute permissions to the same page. A higher-privileged VMPL can use \texttt{RMPADJUST} to adjust the permissions of lower-privileged VMPLs. This mechanism allows a privileged component to protect its own memory and expose selected pages to lower VMPLs with restricted permissions.

\textbf{SVSM}. The SVSM is a privileged service framework for SEV-SNP guests~\cite{svsm_spec}. SVSM runs in VMPL0 and provides security services to a guest OS running at a lower VMPL. The guest OS invokes SVSM services through a GHCB-based calling path. Existing systems use SVSM to host security services such as virtual Trusted Platform Module (vTPM), key management, and kernel integrity protection~\cite{narayanan2023remote,mei2024svsm,mei2025vmpl}. In PS-UIE, the SVSM-hosted vTPM provides virtual Platform Configuration Register (vPCR) and Quote support to bind and verify runtime evidence.

\subsubsection{Linux Execution Paths and Integrity Measurement}

\textbf{Linux User-Space Execution Paths}. Linux grants execute permission to user-space memory mainly through three common paths. First, the \texttt{execve} path loads a new program and, for ELF binaries, the program interpreter; executable \texttt{PT\_LOAD} segments represent the executable content. Second, the \texttt{mmap(PROT\_EXEC)} path creates executable mappings, including file-backed mappings for dynamically loaded shared objects (DSOs). Third, the \texttt{mprotect}-based Non-eXecutable-to-eXecutable (NX$\rightarrow$X) path upgrades an existing non-executable mapping to executable. These paths determine when file-backed user-space executable objects become executable, while anonymous executable memory has no backing file and is commonly associated with Just-In-Time (JIT)-style code generation or fileless code injection. Prior work treats \texttt{execve} as an execution-related sensitive system call and \texttt{mmap}/\texttt{mprotect} as memory-permission system calls that remain necessary for program and library loading~\cite{jelesnianski2023protect}.

\textbf{Integrity Measurement and Enforcement}. Integrity measurement computes a cryptographic digest over a target object or selected executable content~\cite{sailer2004design}. A system can record the measured digest for auditing or compare it with trusted reference values~\cite{sailer2004design,ima_appraisal2012}. Integrity enforcement uses the comparison result to decide whether the measured object should be allowed or denied~\cite{ima_appraisal2012}. Measurement-only mechanisms provide visibility into runtime state, while enforcement mechanisms can block objects that do not match the expected integrity state.

\begin{table*}[!t]
\centering
\caption{Comparison of Representative Related Works and PS-UIE}
\label{tab:rw_compare}
\scriptsize
\renewcommand{\arraystretch}{1.2}
\setlength{\tabcolsep}{4pt}
\newcommand{\cmark}{\checkmark}
\newcommand{\xmark}{\texttimes}
\newcommand{\pmark}{$\triangle$}
\begin{tabular}{r c c c c c c c c}
\toprule
\multirow{2}{*}{\makecell[r]{\textbf{Ref.}}} &
\multicolumn{3}{c}{\textbf{Protection Context}$^{1}$} &
\multicolumn{3}{c}{\textbf{Functions}$^{2}$} &
\multicolumn{2}{c}{\textbf{Evidence}$^{3}$} \\
\cmidrule(lr){2-4}\cmidrule(lr){5-7}\cmidrule(lr){8-9}
&
\textbf{Platform} &
\textbf{Authority} &
\textbf{Main Target} &
\textbf{Measurement} &
\textbf{Policy Ctrl.} &
\textbf{Exec-Permission Ctrl.} &
\textbf{Log} &
\textbf{PCR/vPCR} \\
\midrule

SVSM-vTPM (2023)~\cite{narayanan2023remote}
& CVM & VMPL0 & Service
& \pmark & \xmark & \xmark & \xmark & \cmark \\

Cabin (2024)~\cite{mei2024cabin}
& CVM & VMPL0 & Program
& \xmark & \pmark & \xmark & \xmark & \xmark \\

NestedSGX (2025)~\cite{wang2025road}
& CVM & VMPL0 & Enclave
& \cmark & \xmark & \xmark & \xmark & \xmark \\

\midrule

IMA-Appraisal (2012)~\cite{ima_appraisal2012}
& REE & Linux & File
& \cmark & \cmark & \xmark & \cmark & \cmark \\

TZ-IMA (2022)~\cite{song2022tz}
& Non-CVM TEE & TrustZone & Program
& \cmark & \cmark & \xmark & \cmark & \cmark \\

SMILE (2022)~\cite{zhou2022smile}
& Non-CVM TEE & SGX & Enclave
& \pmark & \xmark & \xmark & \xmark & \xmark \\

Dimac (2024)~\cite{song2024dimac}
& Non-CVM TEE & TrustZone & Container
& \cmark & \xmark & \xmark & \cmark & \cmark \\

\midrule

VMPL-KMI (2025)~\cite{mei2025vmpl}
& CVM & VMPL0 & LKM
& \cmark & \xmark & \xmark & \cmark & \xmark \\

\midrule

\textbf{PS-UIE (Ours)}
& \textbf{CVM} & \textbf{VMPL0} & \textbf{User-Space Exec. Obj.}
& \textbf{\cmark} & \textbf{\cmark} & \textbf{\cmark}
& \textbf{\cmark} & \textbf{\cmark} \\

\bottomrule
\end{tabular}

\vspace{0.5ex}
\begin{flushleft}
\footnotesize
\textbf{Notes:} 
\cmark~indicates supported, \pmark~indicates partially supported or supported for a different target, and \xmark~indicates not supported.
$^{1}$\textbf{Protection Context} indicates the platform, privileged authority, and main protected target of each work.
$^{2}$\textbf{Functions} indicates whether the work supports runtime measurement, policy-based control, and execute-permission control.
$^{3}$\textbf{Evidence} indicates whether the work provides exported runtime logs and PCR-like or vPCR-based evidence binding.
\end{flushleft}
\vspace{-8pt}
\end{table*}

\subsection{Related Work}
\label{sec:related}

This section reviews work related to PS-UIE and compares representative systems with PS-UIE in Table~\ref{tab:rw_compare}.

\subsubsection{Trusted Services inside CVMs}

Recent works use privileged in-CVM service layers or CVM-supported isolation to provide security functions that do not rely on the untrusted hypervisor. SVSM-vTPM~\cite{narayanan2023remote} provides a separate vTPM instance in the SVSM for each CVM, and SVSM-KMS~\cite{mei2024svsm} protects key-management logic in the SVSM. VeriSMo~\cite{zhou2024verismo} studies a verified security module for CVMs, while CPC~\cite{chen2024cpc} supports secure execution of CVM maintenance modules. NestedSGX~\cite{wang2025road}, Veil~\cite{ahmad2023veil}, Cabin~\cite{mei2024cabin}, 00SEVen~\cite{schwarz202400seven}, WALLET~\cite{sabanic2025confidential}, and Shelter~\cite{zhang2025complementing} further explore enclave-like execution, protected services, untrusted program confinement, privileged in-VM introspection, confidential serverless execution, and application-level confidential execution, respectively. TETD~\cite{wang2025tetd} provides trusted execution inside a TDX Trust Domain (TD) without relying on in-VM privilege layering.

PS-UIE follows the direction of privileged in-CVM protection, but its goal differs from these approaches. Existing work mainly provides secure services, supports CVM maintenance, or builds isolated execution domains. PS-UIE instead focuses on runtime integrity inside CVMs by measuring and controlling execute-permission grants for user-space executable objects.

\subsubsection{Runtime Integrity Measurement and Enforcement}

Runtime integrity measurement and enforcement have been widely studied in Rich Execution Environments (REEs), non-CVM TEEs, and CVMs. Linux IMA and its extensions~\cite{sailer2004design,ima_appraisal2012,luo2019container,eckel2021userspace} measure or appraise targets inside REEs. These approaches improve software, file, and container integrity in REEs, but their measurement path and evidence state still rely on the REE Trusted Computing Base (TCB). TrustZone-assisted systems measure REE kernels, applications, containers, or hybrid platforms from the Secure World~\cite{kanonov2016secure,dong2020kims,song2022tz,song2024dimac,zhao2024trusted,liu2023tzeamm,mao2025tpm2,wang2024towards}. Other TEE-assisted systems measure runtime state for enclaves or VMs~\cite{zhou2022smile,ozga2021triglav}. OPTEE-RA~\cite{suzaki2025openssf} and PDRIMA~\cite{mao2025pdrima} further study TEE-side measurement and attestation for TrustZone-based TEEs.

Work on CVM-based integrity measurement is more closely related to PS-UIE. VMPL-KMI~\cite{mei2025vmpl} uses SVSM/VMPL0 to protect the integrity of Linux Kernel Modules (LKMs) inside CVMs. It measures LKMs during loading and protects their code and read-only data pages with VMPL-based memory permissions. This design reduces the trust placed in Linux/VMPL1 for LKM integrity. However, VMPL-KMI focuses only on kernel modules and does not target user-space executable objects.

%% file: tex/3-system.tex
\section{System Overview}
\label{sec:system}

This section presents the threat model, security requirements, and system architecture of PS-UIE.

\subsection{Threat Model}
\label{sec:threat_model}

We consider three categories of threats that define the security scope of PS-UIE:

1) \textbf{Threats from outside the CVM:} 
An attacker may control the host environment, including the hypervisor, host OS, and external software stack, to observe exposed interfaces and replay, inject, or modify messages that cross the CVM boundary. The attacker may also supply tampered executable objects before deployment~\cite{zhou2024verismo}.

2) \textbf{Threats at user-space execution boundaries:} 
An attacker may abuse legitimate execution or memory-management interfaces to make unauthorized user-space code executable~\cite{jelesnianski2023protect}. The attacker may also attempt to introduce anonymous executable memory, for example, through fileless code injection or JIT-generated code.

3) \textbf{Threats against evidence:} 
An attacker may control the network path between the CVM and the verifier, replay evidence, tamper with exported logs, or force the CVM to operate under a stale policy state to hide the current runtime integrity state~\cite{narayanan2023remote}.

\subsection{Security Requirements}
Based on the threat model, we derive the following security requirements (\textbf{SR}s) for PS-UIE.

\textbf{SR1: Privilege-Separated Trust Anchor.}
The integrity authority must be separated from the measured user-space executable objects hosted by the guest OS by residing in a higher-privileged protected domain, such as VMPL0.

\textbf{SR2: Mandatory Checking of Covered Execute-Permission Grant Paths.}
The system must ensure that each event on a covered execute-permission grant path is checked before execute permission is granted.

\textbf{SR3: Non-Granting Fast Path Delegation.}
If a guest-side fast path is used, it must act only as a conservative filter and must never grant execute permission.

\textbf{SR4: Authentic and Fresh Policy State.}
The system must ensure that monitor-side policy state is authentic, fresh, and resistant to tampering, replay, and rollback.

\textbf{SR5: Tamper-Evident Evidence and Verifiable Replay.}
The system must record security-critical runtime events as tamper-evident evidence and bind the evidence state to a trust anchor, so that a verifier can replay and validate the claimed runtime integrity state.

\subsection{Architecture Description}
\label{sec:arch_description}

\begin{figure*}[!htbp]
    \centering
    \includegraphics[width=1\textwidth,trim=18 20 35 18,clip]{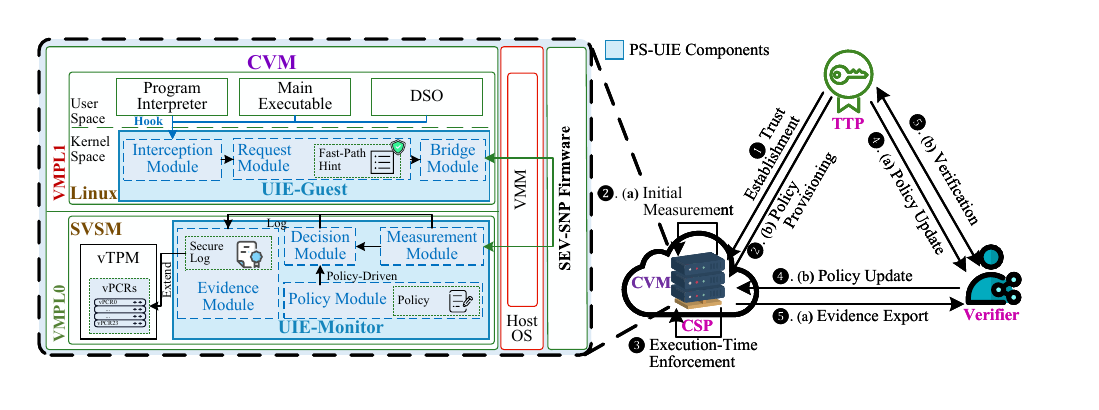} 
    \caption{Architecture and workflow of PS-UIE. The left side shows the two in-CVM components, UIE-Guest in Linux/VMPL1 and UIE-Monitor in SVSM/VMPL0. The right side shows the workflow among the TTP, CSP-hosted CVM, and verifier.}
    \label{fig:description}
\end{figure*}

Fig.~\ref{fig:description} illustrates the architecture of PS-UIE, which consists of three participants and two in-CVM components.

\textbf{Participants.} Denoted by purple letters in Fig.~\ref{fig:description}:

1) \textbf{Trusted Third Party (TTP)}. The TTP is the trusted authority of the system. It prepares trusted deployment materials. It also maintains the reference values used by the verifier to assess the integrity state of the CVM.

2) \textbf{Cloud Service Provider (CSP)}. The CSP provides cloud infrastructure and hosts the SEV-SNP-based CVM for tenants.
    
3) \textbf{Verifier}. The verifier is the tenant-side challenger. It issues fresh challenges, validates attestation materials and exported runtime evidence.

\textbf{Components.} The blue dashed box in Fig.~\ref{fig:description} shows the two in-CVM components inside the CSP-hosted CVM:

1) \textbf{UIE-Guest}. UIE-Guest runs in the Linux/VMPL1 kernel as the guest-side event collector and request forwarder. It contains an \textit{Interception Module} for capturing covered execute-permission grant events, a \textit{Request Module} for constructing normalized measurement and enforcement requests, and a \textit{Bridge Module} for relaying requests, policy inputs, and evidence export interactions to UIE-Monitor. UIE-Guest does not make final integrity decisions.

2) \textbf{UIE-Monitor}. UIE-Monitor runs in SVSM/VMPL0 and acts as the higher-privileged integrity authority. It contains a \textit{Policy Module}, a \textit{Measurement Module}, a \textit{Decision Module}, and an \textit{Evidence Module}. These modules maintain monitor-side policy state, measure covered user-space executable objects, produce final allow/deny decisions, and maintain a Secure Log (SL) for runtime evidence.

\textbf{PS-UIE Workflow}. Fig.~\ref{fig:description} illustrates the workflow supported by the \emph{participants} and \emph{components}:

1) \emph{Trust Establishment} (Phase \circnum{1}). The TTP prepares trusted deployment materials and bootable image components and securely delivers them to the CSP~\cite{mao2025towards}. The CSP then uses them to launch the CVM, and UIE-Monitor becomes active in SVSM/VMPL0.

2) \emph{Initial Measurement and Policy Provisioning} (Phase \circnum{2}). UIE-Monitor bootstraps policy state using initial runtime measurements and a TTP-endorsed policy package.

3) \emph{Execution-Time Enforcement} (Phase \circnum{3}). UIE-Guest intercepts and identifies covered execute-permission grant events. It forwards the corresponding requests to UIE-Monitor for measurement and policy-based decisions. If a conservative fast path is enabled, UIE-Guest may reject definite misses early.

4) \emph{Policy Update} (Phase \circnum{4}). This phase is optional and is used when the monitor-side policy needs to evolve at runtime. A new TTP-endorsed policy package is delivered to the CVM. UIE-Monitor validates the package, rejects stale or rolled-back updates, and switches to the new policy state.

5) \emph{Evidence Export and Verification} (Phase \circnum{5}). UIE-Monitor records security-critical runtime events as verifiable evidence. The verifier obtains and replays the evidence to validate the claimed runtime integrity state of the CVM.

%% file: tex/4-design.tex
\section{System Design}
\label{sec:design}

This section presents the design of PS-UIE by following the core phases of the workflow in Fig.~\ref{fig:description}. Phase~\circnum{1} establishes the initial trust anchor and is introduced in Section~\ref{sec:arch_description}. We treat it as the prerequisite for the runtime integrity workflow. Therefore, this section focuses on Phases~\circnum{2}--\circnum{5}, which form the main runtime integrity workflow of PS-UIE. Together, these phases realize the three mechanisms: \textit{policy lifecycle management} in Phases~\circnum{2} and \circnum{4}, \textit{execution-time integrity enforcement} in Phase~\circnum{3}, and \textit{evidence export and verification} in Phase~\circnum{5}.

\subsection{Policy Lifecycle Management}
\label{sec:policy_lifecycle}

The Policy Module manages the monitor-side policy state used by UIE-Monitor. In PS-UIE, the TTP endorses policy packages, UIE-Guest in Linux/VMPL1 forwards them, and UIE-Monitor validates and installs them in VMPL0/SVSM.

\subsubsection{Policy Model and Package}

PS-UIE uses a digest-based policy model. A policy contains a system mode, a default action, and a set of allow and deny rules. Each rule specifies an execution event, an object role, an expected object digest, and a decision. A Policy Rule (PR) can be described as:

\[
    PR = \langle event, role, digest, decision \rangle.
\]

For each covered request, UIE-Monitor measures the target and matches the tuple \((event, role, digest)\) against the installed policy. If no rule matches, UIE-Monitor applies the default action.

A policy is delivered to UIE-Monitor as a signed policy package. The package contains a compiled binary policy table and secure-installation metadata. The metadata includes the package version, signature algorithm, policy epoch, and an expected vPCR[16] value. The epoch prevents stale or rolled-back policy installation. The expected vPCR[16] value can bind the package to the runtime state measured before installation. UIE-Monitor verifies the package before it becomes the active monitor-side policy state.

If the policy enables the fast path, the signed package carries the corresponding fast path hint. UIE-Monitor validates the package and publishes the hint to UIE-Guest. The use of this hint is described in Section~\ref{sec:fast_path}.

\subsubsection{First Policy Provisioning}

First policy provisioning is the core task in Phase~\circnum{2}. Before the first policy is installed, UIE-Monitor runs in a bootstrap measurement stage and records the initial measurements of covered user-space executable objects. This step helps extend trust from the launch state to the pre-enforcement user-space runtime state by allowing the TTP to check the early user-space execution state.

The TTP obtains the report described in Section~\ref{sec:evidence_export}, which includes the initial measurements from UIE-Monitor. After validating the report against its reference values, the TTP signs an initial policy package and binds it to the validated vPCR[16] value. The package is delivered to the CVM. UIE-Monitor then verifies the signature, epoch, and vPCR[16] binding before installing it.

If the package is accepted, PS-UIE moves from bootstrap measurement to policy-based enforcement. If the package is rejected, UIE-Monitor leaves the monitor-side state unchanged and does not install the rejected package.

\subsection{Execution-Time Integrity Enforcement}
\label{sec:execution_enforcement}

\textit{Execution-time integrity enforcement} is the core task in Phase~\circnum{3}. After policy installation, UIE-Guest forwards covered execute-permission grant events to UIE-Monitor, which measures the target object, records evidence, and returns the final decision.

\subsubsection{Executable Objects and Boundaries}

This subsection defines the executable objects and enforcement boundaries covered by PS-UIE.

\textbf{Covered Executable Objects}. PS-UIE covers user-space executable objects in Linux, including \textit{program interpreters}, \textit{main executables}, and \textit{DSOs}. These objects correspond to ELF interpreters such as the dynamic loader, main executable images loaded by Linux, and executable shared objects mapped from regular files.

\textbf{Enforcement Boundaries}. PS-UIE defines execute-permission grant boundaries as its enforcement boundaries. PS-UIE covers three paths: \texttt{execve}, \texttt{mmap(PROT\_EXEC)}, and \texttt{mprotect(NX$\rightarrow$X)}. UIE-Guest hooks these paths before Linux installs executable mappings. The \texttt{execve} path covers the main executable and the program interpreter before control enters user space. The \texttt{mmap(PROT\_EXEC)} path covers file-backed executable mappings, including DSOs loaded at runtime. The \texttt{mprotect(NX$\rightarrow$X)} path covers file-backed mappings first created as non-executable and later upgraded to executable.

In the current design, PS-UIE does not treat anonymous executable memory as a covered file-backed executable object. If an anonymous mapping attempts an NX$\rightarrow$X transition, PS-UIE applies strict default-deny behavior. This design keeps the enforcement scope focused on file-backed user-space executable objects.

\subsubsection{Request and Measurement}

This subsection describes how UIE-Guest builds a measurement request from an intercepted execution event and how UIE-Monitor produces the object digest for policy decisions.

\textbf{Request Construction}. At each execute-permission grant boundary, UIE-Guest collects the event type, object role, and file identity to describe the target object. These fields identify the permission-grant path, the target file-backed object, and the executable bytes to be measured. UIE-Guest also attaches auxiliary audit information, such as the virtual address range and file offset. UIE-Guest builds a normalized request and sends it to UIE-Monitor through the Bridge Module. The Bridge Module delivers the request as an SVSM service call through the standard SVSM/GHCB calling path, and the Linux/VMPL1 kernel triggers a VMGEXIT-based transition to enter SVSM/VMPL0.

\textbf{Object Measurement}. UIE-Monitor receives the request in SVSM/VMPL0 and measures the executable content described by the request. For a main executable or a program interpreter, UIE-Monitor measures the executable \texttt{PT\_LOAD} segments of the corresponding ELF object. For a file-backed shared object, UIE-Monitor measures the covered executable ranges of the object. UIE-Monitor reads the corresponding guest physical memory ranges and computes the digest over the selected executable ranges. This process produces an object-level digest.

\subsubsection{Decision and Grant}

This subsection describes how UIE-Monitor makes a decision and how UIE-Guest applies this decision at the execute-permission grant boundary.

\textbf{Policy-based Decision}. After the Measurement Module produces the object digest, the tuple \((event, role, digest)\) is passed to the Policy Module. The Policy Module checks deny rules, allow rules, and the default action before returning the decision. In \texttt{MEASURE} mode, UIE-Monitor records the event without blocking execution based on the policy result. In \texttt{ENFORCE} mode, the returned result becomes the final decision.

\textbf{Permission Grant}. UIE-Guest applies the decision returned by UIE-Monitor at the original enforcement boundary. The decision is returned to Linux, and the decision alone determines whether execute permission is granted. If the decision is \texttt{ALLOW}, UIE-Guest lets the corresponding execution path continue so that Linux can grant execute permission to the target object. If the decision is \texttt{DENY}, UIE-Guest aborts the permission grant and reports a policy rejection to the caller.

\subsection{Evidence Export and Verification}
\label{sec:evidence_export}

\textit{Evidence export and verification} is the core task of Phase~\circnum{5}. During execution-time enforcement, UIE-Monitor records security-critical runtime events as verifiable runtime evidence. This section describes the evidence structure and export procedure.

\subsubsection{Evidence Structure}

The Evidence Module maintains an SL in SVSM/VMPL0 as an append-only log and binds it to vPCR[16] of the vTPM in SVSM~\cite{narayanan2023remote}. Linux only relays the exported log for verification.

\textbf{Secure Log}. The SL uses the fixed record format. Each Log Record (LR) describes one security-critical event, such as policy installation, policy update, execution measurement, or policy-based decision:
\[
    LR = \langle EM, OC, ER, LRH \rangle,
\]
where \(EM\) is event metadata, \(OC\) is object context, \(ER\) is the event result, and \(LRH\) is the log record hash. The \(EM\) fields include the record number, vPCR index, event type, and policy epoch. For execution-related records, the \(OC\) fields include the object role, file identity, executable range, and object digest. The \(ER\) fields include the policy mode, decision, reason code, and digest scope.

The \(LRH\) commits to the record content:
\[
    LRH = Hash(EM \parallel OC \parallel ER).
\]
UIE-Monitor computes \(LRH\) over a canonical binary encoding of these fields, so an external verifier can recompute the same hash during replay and detect record modification.

\textbf{vPCR Binding}. After UIE-Monitor creates a log record, it extends the record hash \(LRH\) into vPCR[16] of the vTPM used by PS-UIE:
\[
    vPCR[16] \leftarrow Hash(vPCR[16] \parallel LRH).
\]
The SL stores the runtime context needed for verification. The vPCR[16] stores a compact commitment to this ordered log state. If an attacker deletes, modifies, reorders, or forges SL entries after export, replaying the SL produces a vPCR[16] value that differs from the quoted vPCR state. Thus, PS-UIE provides tamper-evident runtime evidence.

\subsubsection{Evidence Export}

The evidence export procedure lets a verifier obtain a fresh, integrity-protected snapshot of the evidence maintained by UIE-Monitor. The procedure follows a challenge-response workflow. UIE-Guest only relays the request and response. UIE-Monitor and the vTPM produce the evidence state and the Quote inside SVSM/VMPL0.

\textbf{Cryptographic Materials}. The export procedure uses the following materials:

1) $Nonce$. The verifier generates a fresh nonce as the challenge for each evidence request.

2) $PK_{AK}$/$SK_{AK}$. This key pair is the vTPM attestation key pair in SVSM. The vTPM uses \(SK_{AK}\) to sign the Quote. The verifier uses \(PK_{AK}\) to validate the Quote.

\textbf{Export Workflow}. The export workflow consists of four steps:

1) Step~\circled{1}. The verifier sends a request with a fresh $Nonce$ to the CVM. UIE-Guest receives the request and forwards it to UIE-Monitor through the Bridge Module.

2) Step~\circled{2}. The Evidence Module generates a snapshot of the runtime evidence. The exported Integrity Evidence (IE) contains the SL, policy metadata \(PM\), and the current vPCR[16] value. The \(PM\) includes the active policy epoch and policy-related identifiers:
\begin{equation}
    IE \leftarrow \{ SL \parallel PM \parallel vPCR[16] \}.
\end{equation}

3) Step~\circled{3}. UIE-Monitor requests the vTPM to generate a nonce-bound Quote over vPCR[16]. The vTPM signs the Quote with its attestation key:
\begin{equation}
    Quote \leftarrow \operatorname{Quote}_{SK_{AK}}(Nonce, vPCR[16]).
\end{equation}
This Quote protects the freshness and integrity of the reported vPCR state. The verifier later checks the integrity of the SL by replaying it and comparing the recomputed vPCR value with the quoted vPCR value.

4) Step~\circled{4}. UIE-Monitor packages the evidence response as Attestation Evidence (AE):
\begin{equation}
    AE \leftarrow \{ Nonce \parallel IE \parallel Quote \}.
\end{equation}
UIE-Guest then relays \(AE\) to the verifier. Thus, the evidence export procedure is complete.

%% file: tex/5-implementation.tex
\section{Implementation}
\label{sec:implementation}

We implemented PS-UIE on an AMD SEV-SNP server by modifying the Linux kernel and Coconut-SVSM~\cite{amdsev_github,coconut_svsm_github}. The prototype includes about 4,100 lines of code (LoC) for UIE-Guest in Linux/VMPL1, 2,500 LoC for UIE-Monitor in Coconut-SVSM/VMPL0, and 1,450 LoC for policy packaging, evidence export, and verifier-side replay tools.

%% file: tex/6-evaluation.tex
\section{Evaluation}
\label{sec:evaluation}

This section presents PS-UIE's security analysis and performance evaluation.

\subsection{Security Analysis}
\label{sec:sub_security_analysis}

This subsection analyzes how PS-UIE addresses \textbf{SR1}--\textbf{SR5} under the threat model in Section~\ref{sec:threat_model}.

\subsubsection{Trust Anchor and Policy Integrity -- SR1, SR4}

PS-UIE addresses \textbf{SR1} by separating the integrity authority from Linux/VMPL1 and placing it in UIE-Monitor, which runs in SVSM/VMPL0. UIE-Monitor maintains policy state, performs measurement, and produces final decisions. Linux/VMPL1 only intercepts events and forwards requests, without serving as the authority for object integrity or execution authorization. Section~\ref{sec:discussion} discusses the security implications and possible hardening of this design. Under our threat model, PS-UIE assumes that the covered hooks faithfully issue requests to VMPL0, but UIE-Monitor does not trust guest-side integrity decisions.

PS-UIE addresses \textbf{SR4} through signed policy packages and monitor-side validation. The TTP endorses policy materials before delivery. UIE-Monitor verifies the signature, policy epoch, and vPCR[16] binding before installing or updating a policy. The epoch prevents rollback. The vPCR[16] binding prevents policy installation under an unexpected state. Because Linux/VMPL1 only forwards policy inputs, forged or modified packages cannot cause UIE-Monitor to install an invalid policy.

\begin{figure}[!t]
\centering
\includegraphics[width=3.5in,trim=5 7 5 7,clip]{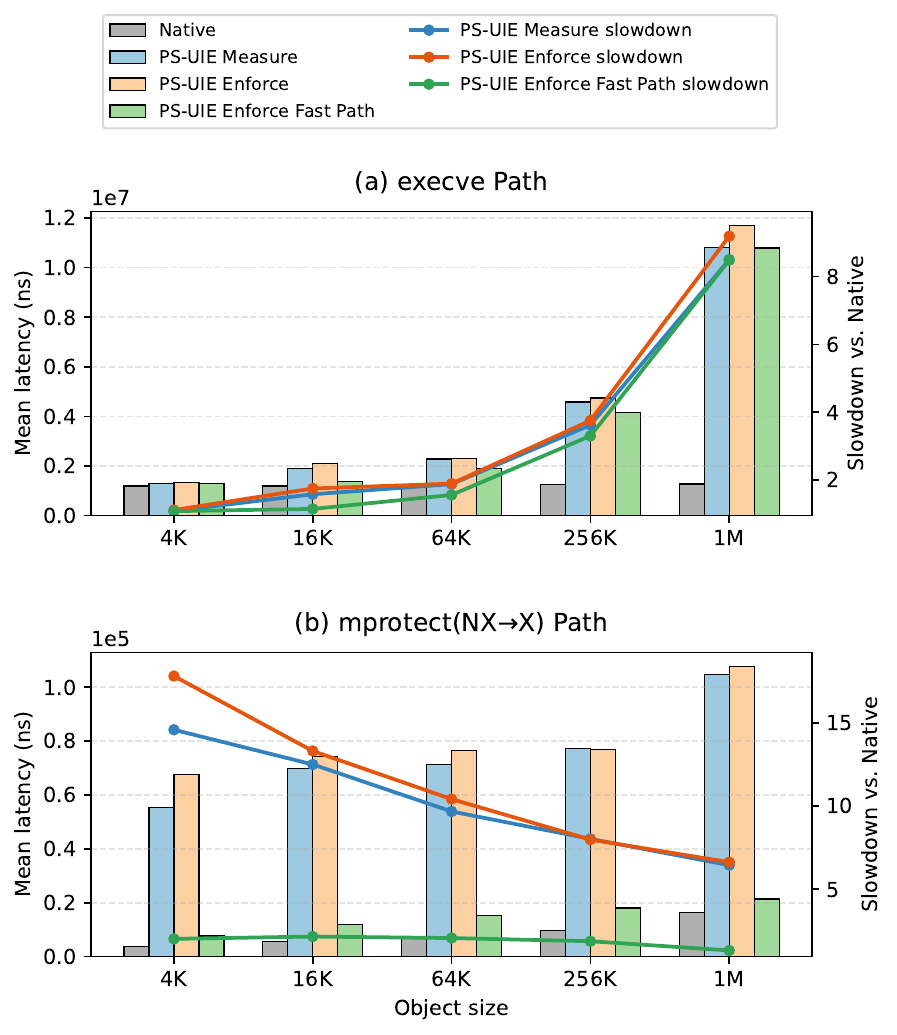}
\caption{Overhead of the \texttt{execve} and \texttt{mprotect(NX$\rightarrow$X)} paths. Bars show mean latency and lines show slowdown over \textit{Native}.}
\label{fig:execve_mprotect}
\vspace{-10pt}
\end{figure}

\subsubsection{Execution-Time Enforcement Security -- SR2, SR3}

PS-UIE addresses \textbf{SR2} by enforcing integrity checks on covered execute-permission grants before covered user-space executable objects become executable. PS-UIE places enforcement boundaries on the Linux paths that grant execute permission. Thus, a covered file-backed object must be measured and allowed by UIE-Monitor before execution continues. UIE-Monitor performs the measurement in VMPL0 and evaluates the policy using the tuple \((event, role, digest)\). Under our threat model, using another covered grant path or replacing the file is insufficient to authorize execution, because the decision depends on monitor-side measurement and decision.

PS-UIE also prevents anonymous executable memory from being accepted as a covered file-backed object. When an anonymous mapping attempts an NX$\rightarrow$X transition, PS-UIE applies the default-deny baseline. This default-deny behavior prevents anonymous executable memory from being accepted under the stated policy scope.

PS-UIE addresses \textbf{SR3} by keeping the fast path conservative. The Bloom-filter hint can only reject definite misses and cannot grant execute permission. Because UIE-Monitor publishes the hint only after policy validation and exposes it to VMPL1 with read-only permission, a missing, invalid, or corrupted hint cannot create an unauthorized allow decision. The allow boundary remains in VMPL0/SVSM.

\subsubsection{Evidence Security and Verifiable Replay -- SR5}

PS-UIE addresses \textbf{SR5} by maintaining the SL in UIE-Monitor in SVSM/VMPL0 and binding the evidence state to vPCR[16]. Since UIE-Monitor extends each \(LRH\) into vPCR[16], any deletion, modification, reordering, or forgery of exported records causes verifier-side replay to diverge from the nonce-bound quoted vPCR[16] value. The nonce-bound Quote prevents stale evidence from being accepted as fresh. The verifier further checks object digests against TTP-endorsed reference values and verifies recorded decisions against the active policy. Thus, PS-UIE provides tamper-evident and verifiable runtime evidence without trusting Linux/VMPL1 or the network path.

\begin{figure*}[t]
\centering
\includegraphics[width=0.8\textwidth,trim=5 7 5 7,clip]{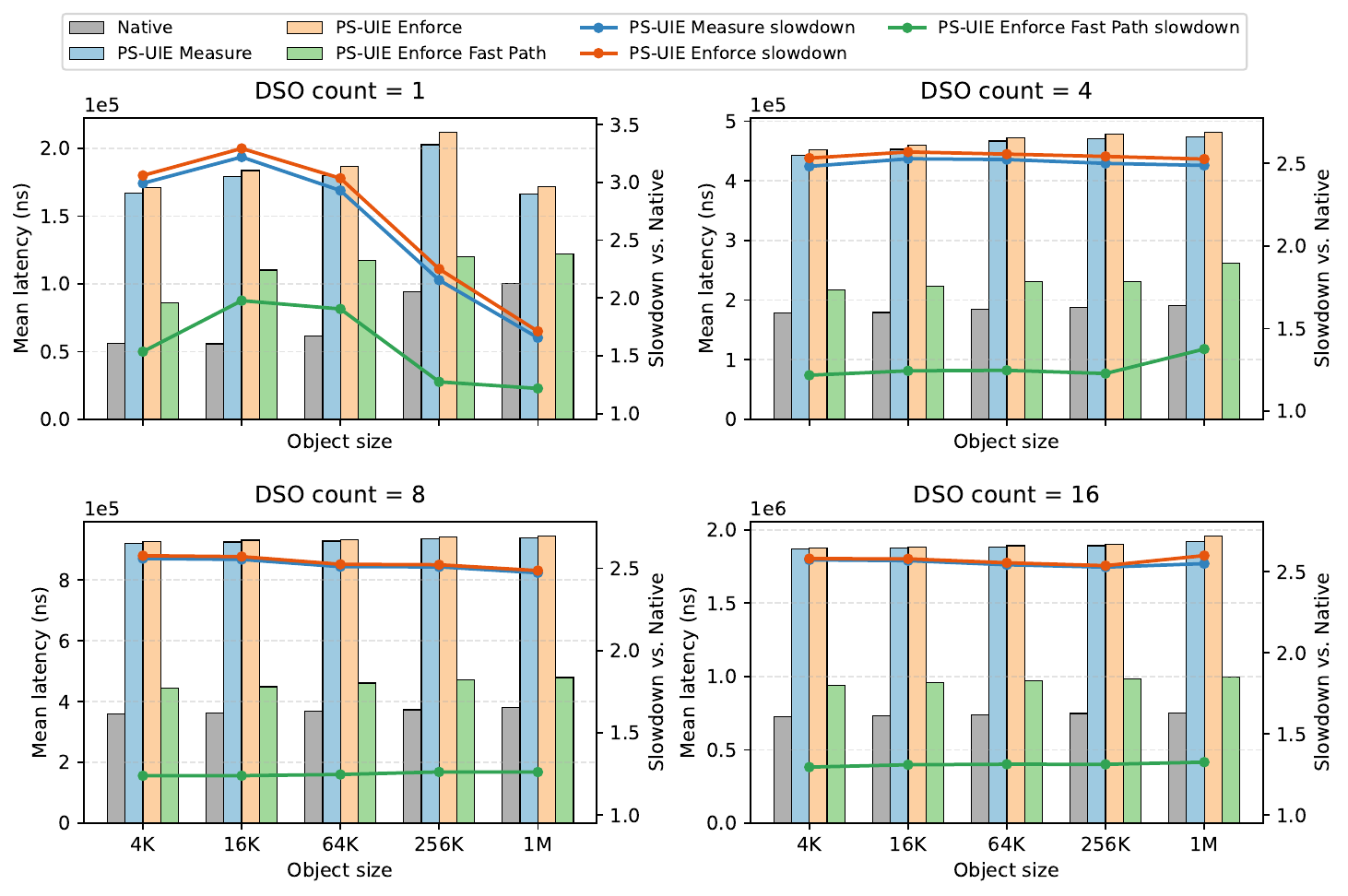}
\caption{Overhead of the \texttt{mmap(PROT\_EXEC)} path with file-backed DSOs under different DSO counts. Bars show mean latency and lines show slowdown over \textit{Native}.}
\label{fig:dso_overhead}
\vspace{-10pt}
\end{figure*}

\subsection{Performance Evaluation}
\label{sec:sub_performance}

This subsection evaluates PS-UIE's performance on execution paths.

\subsubsection{Experimental Setup}

All experiments run on an AMD SEV-SNP server equipped with an AMD EPYC 7763 64-core processor and 256~GB of RAM. The software stack follows the implementation in Section~\ref{sec:implementation}. The host runs Ubuntu 22.04 with Linux kernel v6.1.0 and QEMU v7.2.0. The guest runs Ubuntu 20.04 with Linux kernel v6.5.0. Coconut-SVSM version v2026.01-devel-42 runs in VMPL0 in the guest.

We compare four configurations:

1) \textbf{\textit{Native}} disables PS-UIE.

2) \textbf{\textit{PS-UIE Measure (Measure)}} enables measurement and evidence recording without policy enforcement.

3) \textbf{\textit{PS-UIE Enforce (Enforce)}} adds policy enforcement to \textit{PS-UIE Measure}.

4) \textbf{\textit{PS-UIE Enforce with Fast Path (Enforce FP)}} further enables the Bloom-filter hint-based fast path.

We report mean latency, execution time, rating, or throughput according to the benchmark metric. We also report the normalized slowdown or ratio over \textit{Native}. For \texttt{Redis}, we report startup latency and request throughput.

\subsubsection{Execution-Path Overhead}

This experiment evaluates PS-UIE overhead on execute-permission grant paths. Fig.~\ref{fig:execve_mprotect} reports results for \texttt{execve} and \texttt{mprotect(NX$\rightarrow$X)}, and Fig.~\ref{fig:dso_overhead} reports results for \texttt{mmap(PROT\_EXEC)} with file-backed DSOs. We vary executable object size from 4~KB to 1~MB and report mean latency and slowdown over \textit{Native}.

\textbf{\texttt{execve} Path.} PS-UIE introduces modest overhead for small objects and higher overhead for large objects. At 4~KB, \textit{Measure}, \textit{Enforce}, and \textit{Enforce FP} incur 1.08$\times$, 1.11$\times$, and 1.08$\times$ slowdown, respectively. At 1~MB, the slowdowns increase to 8.50$\times$, 9.19$\times$, and 8.49$\times$. This trend shows that \texttt{execve} overhead is mainly determined by the amount of executable content measured before execution starts.

\textbf{\texttt{mprotect(NX$\rightarrow$X)} Path.} \textit{Measure} and \textit{Enforce} incur higher overhead than \textit{Native} because PS-UIE checks the permission upgrade before granting execute permission. The fast path configuration reduces the slowdown to 1.32--2.15$\times$ by rejecting definite-miss requests before entering UIE-Monitor. This result shows that the Bloom-filter hint reduces unnecessary monitor transitions.

\textbf{\texttt{mmap(PROT\_EXEC)} Path.} The overhead increases with the number of DSOs because each file-backed executable mapping may trigger a separate enforcement request. With one DSO, \textit{Enforce} incurs 1.71--3.29$\times$ slowdown, while \textit{Enforce FP} reduces the slowdown to 1.22--1.98$\times$. With 16 DSOs, \textit{Enforce} incurs 2.54--2.60$\times$ slowdown, while \textit{Enforce FP} reduces the slowdown to 1.30--1.33$\times$. These results show that DSO loading accumulates cost across executable mappings, and that the fast path reduces this cost for definite miss requests.

%% file: tex/8-conclusion.tex
\section{Conclusion}
\label{sec:conclusion}
This paper presented PS-UIE, a privilege-separated approach for enforcing user-space executable integrity inside CVMs. PS-UIE uses a privilege-separated architecture that separates the final integrity authority from the measured user-space executable objects by placing it in a higher-privileged protected domain. Built on this architecture, PS-UIE provides three mechanisms: \textit{policy lifecycle management}, \textit{execution-time integrity enforcement}, and \textit{evidence export and verification}. PS-UIE measures and enforces the integrity of user-space executable objects at execute-permission grant boundaries and records security-critical runtime events as verifiable runtime evidence. We implemented PS-UIE on an AMD SEV-SNP platform by modifying the Linux kernel and the SVSM. The security analysis and performance evaluation show that PS-UIE enforces user-space executable integrity on the covered execute-permission grant paths, provides verifiable runtime evidence, and incurs acceptable overhead on the evaluated workloads.

%% file: tex/9-references.tex
\FloatBarrier
\bibliographystyle{jabbrv_IEEEtran}
\bibliography{main}